\documentclass[12pt]{article}
\usepackage{amsmath}
\usepackage{amssymb}
\usepackage{epsfig}
\usepackage{textcomp}

\newcommand{\alg}{\Sigma}

\newcommand{\cA}{\mathcal{A}}
\newcommand{\cB}{\mathcal{B}}
\newcommand{\cC}{\mathcal{C}}
\newcommand{\cD}{\mathcal{D}}
\newcommand{\cM}{\mathcal{M}}
\newcommand{\cP}{\mathcal{P}}

\newcommand{\cS}{\mathcal{S}}
\newcommand{\cX}{\mathcal{X}}
\newcommand{\cY}{\mathcal{Y}}

\newcommand{\Om}{\Omega}

 at10pt

\begin{document}
\title {Non-separability does not relieve the problem of Bell's theorem}
 \author{Joe Henson\footnote{j.henson@imperial.ac.uk} } \maketitle
\begin{abstract}
This paper addresses arguments that ``separability'' is an assumption of Bell's theorem, and that abandoning this assumption in our interpretation of quantum mechanics (a position sometimes referred to as ``holism'') will allow us to restore a satisfying locality principle.  Separability here means that all events associated to the union of some set of disjoint regions are combinations of events associated to each region taken separately.

In this article, it is shown that: (\textit{a}) localised events can be consistently defined without implying separability; (\textit{b}) the definition of Bell's locality condition does not rely on separability in any way; (\textit{c}) the proof of Bell's theorem does not use separability as an assumption.  If,  inspired by considerations of non-separability, the assumptions of Bell's theorem are weakened, what remains no longer embodies the locality principle.  Teller's argument for ``relational holism'' and  Howard's arguments concerning separability are criticised in the light of these results.  Howard's claim that Einstein grounded his arguments on  the incompleteness of QM with a separability assumption is also challenged.  Instead, Einstein is better interpreted as referring merely  to the existence of localised events.  Finally, it is argued that Bell rejected the idea that separability is an assumption of his theorem.
\end{abstract}
\vskip 1cm

\section{Introduction}

There are a plethora of views about Bell's theorem \cite{Bell:2004}: what its essential meaning is, whether it is a problem, and if so, whether this problem can be solved (see \textit{e.g.}~\cite{Redhead:1987, Cushing:1989,Maudlin:2002}).  Bell's own position, set out most clearly in his final exposition of the subject (\cite{Bell:2004}, 1990, pp.232-248), was that the theorem showed a contradiction between quantum mechanics (and, arguably, experiment) and a reasonable formalisation of a ban on superluminal causal influence.  Broadly the same view has recently been championed by Norsen \cite{Norsen:2006a,Norsen:2007,Norsen:2007a,Norsen:2009}.  As has already been noted in several places \cite{Butterfield:1992a,Brown:2002a, Henson:2013a}, Bell's formulation of local causality is a combination of a Reichenbachian Principle of Common Cause (PCC) \cite{Reichenbach:1956a} with the specification that relativistic past regions are the relevant ones for the PCC.

Bell defended the use of this formulation against the operational idea of lack of superluminal signalling, on the grounds that relying on concepts such as measurements, or fields external to the quantum system, is not sufficient to treat the concept of causality.  His arguments have been followed up and expanded on by Maudlin \cite{Maudlin:2002} and Norsen \cite{Norsen:2009}.  Besides this, there are many other grounds on which we might want to preserve a relativistic PCC.  The most basic reason is that doing so may enhance our understanding of present theory, and increase progress toward improved theories.  When we demand that correlations between distant events have a causal explanation (\textit{e.g.}~magnetic effects), we are led to add detail to our theories (\textit{e.g.}~fields).  Similarly, any basic principle can be employed in reformulations of present theories, and in the search for new theories.  Finally, upholding a principle banning superluminal influence would naturally explain the predicted and observed inability to send superluminal signals.

For all these reasons, the claim that locality can be maintained in modern physics, despite Bell's theorem, is an interesting one.  Some commentators have claimed that the problem can be solved by pointing out implicit assumptions of Bell's theorem that are independent of locality, once that concept is properly construed \cite{Jarrett:1984} (\cite{Cushing:1989}, Jarrett pp.~60-79).  At the extreme, Wessels lists a veritable menagerie of conditions, amounting to (at least) seventeen in total, and shows that these imply a Bell's theorem \cite{Wessels:1985} (\cite{Cushing:1989}, Wessels pp.~80-96).  However, if one wants to rely on one's favourite derivation of Bell's theorem for the purposes of this discussion, one needs to show that the assumptions one makes are equivalent to, or weaker than, what is used (explicitly or implicitly) in the standard versions.  After all, if I added the assumption that I live in London to a derivation of Bell's theorem, that would not make it reasonable for a group of angry realists to drive me out of town in the hope of saving locality\footnote{Maudlin has coined the term ``The Fallacy of the Unnecessary Adjective'' to indicate this problem with descriptions of Bell's theorem \cite{maudlin:2010}, citing ``realistic,'' ``hidden-variable'' and ``counterfactually definite'' as examples of unnecessary adjectives that have cropped up between ``local'' and ``theories''.}.

With this in mind we can turn to the supposed assumption that will be the main topic of this paper: ``separability'' \cite {Howard:1985} \cite{Teller:1986} (\cite{Cushing:1989}, Teller pp.~208-223,  Howard pp.224-253).   As Howard puts it, ``we might...~all along have been testing not simply local hidden variable theories, but separable, local hidden variable theories" (\cite{Howard:1985}, p.195).  Among the definitions Howard makes we find that ``separability says that spatially separated systems possess separate real states'' (\cite{Howard:1985} p.~173) and, in a later treatment, that separability ``is a fundamental ontological principle'' which
\begin{quote}
...~asserts that the contents of any two regions of spacetime separated by a nonvanishing spatiotemporal interval constitute separable physical systems, in the sense that (1) each possesses its own, distinct physical state, and (2) the joint state of the two systems is wholly determined by these separate states.
\end{quote}
(\cite{Cushing:1989}, Howard pp.225-226).  In two widely cited papers,  Howard has discussed the principle in the context of Bell's theorem and Einstein's completeness arguments, claiming that locality and QM can both stand when separability is eliminated \cite{Howard:1985} (\cite{Cushing:1989}, Howard pp.224-253).  Teller writes that our bafflement over Bell's result ``arises from an unnoticed and very deeply seated presupposition we make about ontology'' (\cite{Cushing:1989}, Teller p.~208).  This is \textit{particularism,} another version of separability: ``[i]n application to relativistic theories, particularism takes the form of supposing the theory to apply exclusively to spacetime points and their non-relational properties.'' Teller has made a case for dropping this supposed assumption, arguing that locality and the results of QM can both be saved in this way, and naming the converse principle ``relational holism'' \cite{Teller:1986} (\cite{Cushing:1989}, Teller pp. 208-223).

This type of reasoning has been taken up in many places, for example by Shimony  (\cite{Cushing:1989}, Shimony, pp. 25-37), and in Redhead's influential book on quantum nonlocality, in which ``denial of separability'' is claimed to ``block the derivation of the Bell inequality'' (\cite{Redhead:1987}, p.168).  Developing on this theme, Healey has proposed an ``interactive'' interpretation of QM that relies heavily on non-separability, and discussed holism more broadly \cite{Healey:1989,Healey:1991, Healey:1992, Healey:1994}, and Morganti has built on Teller's ideas \cite{Morganti:2009}.  On the other hand, Laudisa and Maudlin \cite{Laudisa:1995,Maudlin:2002} have disputed some of Howard's claims, while Berkowitz \cite{Berkovitz:1998} and Fine and Winsburg  \cite{Winsberg:2003}, followed by Fogel \cite{ Fogel:2007}, have criticised some parts of Howard's argument while accepting others  (see appendix \ref{a:states}).

Following Howard's second definition, separability has commonly been taken to mean that, for a set of disjoint regions, any statement that can be made about events in the union of the regions is in fact a logical combination of statements about events in each separate region, or equivalently the events in the union ``supervene'' on those in each separate region in the terminology of Teller and Healey\cite{Healey:2007}\footnote{Healy compares definitions of separability in which physical processes supervene on properties defined at points and/or arbitrarily small neighbourhoods of those points (\cite{Healey:2007}, p.46).  See also \cite{Butterfield:2005} for a related discussion.  The main conclusions of this article are independent of this distinction.}.  If this was indeed a necessary implicit assumption of Bell's theorem, and independent of locality, abandoning the principle would be an attractive way to avoid the problem of Bell's theorem.  Healey has made an argument that separability should be abandoned in the context of (classical) gauge theories \cite{Healey:2007}.  If this is so, why not pay the same small price in order to save locality in quantum mechanics as well?

Adding to the motivation to carefully consider such options, in recent years there has been increased interest in causal principles and Bell's theorem fueled mainly by the rise of quantum information studies.  The $\Psi$-epistemic position, for example, has been much discussed and debated \cite{Spekkens:2004a, Harrigan:2010,Bartlett:2012,Pusey:2012}.  It is natural in this approach to hope that dropping some deep assumption in Bell's theorem will relieve the tension between locality and QM, analogously to the way that special relativity relieves the tension between relativity of motion and invariance of the speed of light\footnote{This last point derives from comments made by Spekkens in a seminar \cite{Spekkens:talk2012}.}.  This is a good reason to carefully scrutinise the EPR debate and the assumptions of Bell's theorem from this perspective.  In \cite{Harrigan:2010}, which makes much use of references to Howard, Harrigan and Spekkens state the following: ``A necessary component of any sensible notion of locality is separability".  While this is distinct from Howard's claim, it does leave the door open to the possibility that non-separable or ``holistic'' models could somehow  escape from the conceptual problems of Bell's theorem\footnote{Spekkens later endorsed an different view, however, arguing that Bell's theorem does not rely on separability \cite{Spekkens:2012,Spekkens:talk2012}.}.  Similarly, Ghirardi and Grassi state that ``usually, in discussing EPR-like situations one pays little attention to separability, which is in a sense tacitly assumed as a prerequisite of the locality principle'' (\cite{Ghirardi:1994} p.405).

Ambiguity of terms has often been a problem in these discussions, allowing arguments to move from general principles to quite specific statements about Bell's theorem in ways that are not obviously justified.  In this paper, some mathematical formalisation is sought as an antidote to this.  To discuss the claims it is useful to collect together some cursory definitions of the principles that play a part in this debate, to be enlarged upon later.  The first three are adapted from definitions made by Healey \cite{Healey:2007}.

\paragraph{Locality (principle):} There is no superluminal influence.

\paragraph{Local action (principle):} influence between events in separated regions must be mediated by some event(s) in the intervening regions.

\paragraph{Separability (principle):} all events associated to a disjoint union of regions supervene on the events associated to each of the regions in the union.

\paragraph{Principle of common cause:}  If there is a correlation\footnote{As will become clear, ``correlation'' here takes on a specific meaning: a lack of statistical independence of the events given a probability distribution, \textit{i.e.}~in an obvious notation, $P(A) P(B) \neq P(A \cap B)$.  This is distinct from definitions by which the actual occurrence of two events implies a correlation between them, as discussed in the present context in \cite{Maudlin:2002}.  In line with this, ``event'' has the meaning it does in the theory of stochastic processes and \textit{e.g.}~\cite{Butterfield:2005}, that is, it is a proposition that may or may not be true in any possible realisation (or ``history'') of the process, and to which is associated a probability (or probabilities).  That is, the definition of correlation employed here does not refer to actual particular happenings such as the fact that a particular coin flip produced ``heads'', but the theoretically given probabilities of such results.} between two events, then either one of these events has influenced the other, or there exists some common cause of the correlation (associated to the past), such that, once this cause is conditioned on, the correlation disappears.

\paragraph{} Unless noted, these meanings will be used throughout the paper.  This must be kept in mind especially for the slippery word ``locality,'' which is often given different meanings, but (albeit with some reluctance) has been defined here in such a way as to make common sayings like ``Bell's theorem exposes a tension between locality and QM'' valid.

Separability, it will be demonstrated below, has in many places been confounded with a much weaker principle that will be called the principle of localised events.

\paragraph{Localised events (principle):}  All events can be associated to regions in spacetime in a consistent manner.

\paragraph{} In order to formalise some of these principles, a straightforward and minimal framework is introduced in this article, in section \ref{s:framework}.  Using these tools, the principle of localised events is formalised in section \ref{s:locevents}, and is shown not to imply separability even when a sensible structure is imposed on the assignments of regions to events.  It is then shown in section \ref{s:belllocality} that locality can indeed be defined without reference to separability, and, in section \ref{s:eprb}, that Bell's theorem can be derived without any assumption of separability.  In short, it is neither the case that separability is a necessary part of locality, nor that it is an implicit and necessary assumption of the theorem.  Furthermore, in section \ref{s:weakening} it is argued that two attempts to weaken the assumptions of Bell's theorem inspired by relational holism can quickly be seen to be problematic.  After this is established, in section \ref{s:people} these conclusions are compared to those of Teller and Howard (prefaced with some comments on a common ancestor of their arguments from Jarrett), and several important flaws in the arguments are pointed out.  The article ends with some comments on what Einstein and Bell themselves actually wrote about separability in the context of interpretations from Howard, Harrigan and Spekkens.  Appendix \ref{a:nouvelle} treats a variant of Bell's local causality while in appendix \ref{a:states} it is argued that Howard's ``separability of states'' condition is not a good formalisation of his separability principle.

\section{Stochastic processes in spacetime}
\label{s:framework}
In this section a framework for dealing with causal principles will be explained, similar to that given in \cite{Henson:2005wb}.  In essence the framework is that of stochastic processes, with some extra structure to describe where the events occur in spacetime.  The intention is that the additional structure is only what is necessary for this task, and no more.  In a sense, all that is being done here is to formalise some aspects of Bell's idea of ``local beables''.  Attention will be payed to the issues discussed above while making these definitions.  In particular, this will clarify the assumptions necessary to prove a Bell theorem.  Throughout this section, comments on the connections to the literature are kept to a minimum, allowing the argument to stand for itself. Results will be compared to other treatments more fully in the following section.

The first structure that is necessary is a spacetime $\cM$ in which events will be localised.  We may as well take this to be Minkowski space, although any weakly causal Lorentzian manifold (\textit{i.e.}~one such that past and future causal relations are unambiguously defined as a partial order) will do.  Subsets of $\cM$ will be referred to as ``regions'', and calligraphic script will be used to represent them.  Next, we define a set of histories $\Om$, which may, for example, be a set of spacetime field configurations and/or particle trajectories.  ``Events'' are subsets of $\Om$, for instance, in the previous example, that a particle of some type passes through a region.  As usual with stochastic processes, the set of events is closed under an appropriate class of logical operations.  It forms a $\sigma$-algebra $\alg(\cM)$, meaning that that the set of events is closed under complementation and countable unions of its members.  The $\sigma$-algebra property allows us to consistently define a probability measure $\mu$ on the set of events. For countable history spaces this reduces to the more familiar Boolean algebra, and the distinction will not be crucial for the main points below.  

\subsection{Localised events}
\label{s:locevents}

This leaves open the question of how the events relate to the spacetime.  Firstly, the principle of localised events given in the introduction requires that all events have assigned to them some set of regions.  But, one could argue, this on its own does not seem to embody the principle.  It would surely violate the spirit of the principle if, for instance, an event was associated to a region while its complement was not, or if the same event could be assigned to many disjoint regions.  It is necessary to be explicit about these conditions, to see whether they imply anything about separability.

To answer this, we introduce the following definition:

\paragraph{Localised Events:} to every region in spacetime $\cA$ is associated a subalgebra $\Sigma(\cA)$ of $\Sigma(\cM)$, such that the following two conditions hold:
\begin{gather}
\label{e:loc-cap}
\bigcap_i \alg(\cA_i) = \alg(\bigcap_i \cA_i),\\
\label{e:trivial}
\alg(\emptyset)= \{ \emptyset, \Om \},
\end{gather}
where $\{\cA_i\}$ is any countable collection of regions and $\emptyset$ is the empty set.

\paragraph{} Note that the intersection of a countable set of $\sigma$-algebras is also a $\sigma$-algebra.   We will say that an event $A$ is associated to a region $\cA$ if $A \in \Sigma(\cA)$.

The meaning of these expressions is straightforward.  That a whole subalgebra is associated to each region means that, if some set of events is associated to a region, then combinations (meaning logical combinations, if the events are thought of as propositions) of these events will be too.  So for instance if ``a blue ball passes through $\cA$'' and ``a red ball passes through $\cA$'' correspond to events are associated to $\cA$, then  ``a blue ball and a red ball pass through $\cA$'' also is.  But if we stopped there, we would still have the problem of one event being assigned to many disjoint regions.  Equation (\ref{e:loc-cap}) says that events associated to both $\cA$ and $\cB$ must be associated to the intersection of those two regions.  This has the consequence that one event will not be in the algebras associated to two disjoint regions, unless it is one of the ``trivial events'' associated to the empty region, $\emptyset$ or $\Om$  (although a non-trivial event can be in the algebra associated to \textit{one} non-simply connected region).

This last condition has a number of other interesting implications.  Firstly, when this condition is assumed, $\cA \subset \cB$ implies $\alg(\cA) \subseteq \alg(\cB)$, or in other words an event associated to a region $\cA$ is associated to any region that contains $\cA$, which is as it should be.  Secondly, every event $A$ is associated to a unique ``intrinsic'' region that is contained by all the other regions associated to $A$.  Finally, because a union of a set of regions is a superset of every region in the set, 
\begin{equation}
\label{e:loc-cup}
\bigcup_i \alg(\cA_i) \subseteq \alg(\bigcup_i \cA_i).
\end{equation} 
In particular, for any two regions $\cA$ and $\cB$, all the events that belong to $\cA$ and $\cB$  (and because they form a $\sigma$-algebra, combinations of these events) also belong to $\cA \cup \cB$.

Conditions (\ref{e:loc-cap}) and (\ref{e:trivial}) do not enter into the derivation of the Bell inequalities below.  The only important thing about assuming localised events for that argument is simply that all events are associated to some intrinsic region.  However, these conditions are relevant to the larger discussion in another way.  Exhibiting them clearly shows that assuming the localised events principle need not imply separability, even when the association of regions to events is given a reasonable structure.  In this framework, the separability principle above becomes the following:
\paragraph{Separability:} for a countable set of disjoint regions $\{\cA_i\}$,
\begin{equation}
\label{e:sep}
G \bigl( \bigcup_i \alg(\cA_i) \bigr)= \alg(\bigcup_i \cA_i),
\end{equation} 
where $G(X)$ is the $\sigma$-algebra generated by the collection of events $X$.  We will use the term ``non-separable event' to refer to an event that is not a combination of events occurring in a partition of some region to which it is associated.

\paragraph{} This condition agrees with the separability principle, and thus with the various definitions in the literature.  It is important to note that, from the localised events condition, we could only derive (\ref{e:loc-cup}), which says that the algebra of events associated to the union of many regions \textit{contains} all the events belonging to the constituent regions and the combinations thereof.  In contrast, separability says that, for disjoint regions, these events are \textit{all} that is to be found in the union of the regions.

Unlike (\ref{e:loc-cup}), it is easy to see that (\ref{e:sep}) does \textit{not} follow from our definition of localised events.  Consider an example in which $\alg(\cM)=\{\emptyset,\Om,X,\bar{X}\}$, where the bar indicates complimentation, and in which $\cM$ contains a region $\cX$ that itself contains many disjoint regions.  Let the algebra associated to $\cX$, and all regions containing it, equal $\alg(\cM)$, and all others equal the trivial algebra $\{ \emptyset, \Om \}$.  Clearly this meets the conditions for the existence of localised events, and just as clearly it violates separability because $X$ is non-separable.  Such non-separable events can be added to any given stochastic process with localised events in an obvious way, showing that this argument applies to a large and general class of models.    

\subsection{Bell locality and a consistency condition}
\label{s:belllocality}

The definition of localised events, although natural, does not say much about what these events are supposed to represent.  Can we say that these events correspond to propositions about Bell's local beables?  Should we struggle to define intrinsic (which should rule out events associated to $\cA$ corresponding to such statements as ``a soon-to-be-destroyed ball passes through region $\cA$'') and qualitative (which should rule out events associated to $\cA$ corresponding to statements like ``the man passing through region $\cA$ is the King of Sweden'' unless they are equivalent to some qualitative formulation)  physical properties \cite{Healey:1991, Healey:1994}?

The answer to this will be supplied in the form of more formal conditions.  In spirit, the answer is that the formal events of the stochastic process are intended to represent the kinds of events that physics should be primarily concerned with according to Einstein and Bell, at least while spacetime remains fixed (see sections \ref{s:howard}, \ref{s:einstein} and \ref{s:spekkens}).  That is, the events have two conceptual roles which are deeply intertwined.  Firstly, operational events that are observable in a region (whatever we are comfortable including here) must be associated to that region.  Bell says ``[t]he beables must include the settings of switches and knobs on experimental equipment, the currents in coils, and the readings of instruments'' and he goes on to associate such settings with appropriate regions in a straightforward fashion (\cite{Bell:2004}, 1975, p.~52). Secondly, the events can serve as causal factors for other events, operational or not, allowing causal explanations.  With these rules in place the events become recognisably physical, answering questions about ``beables''.  Questions of whether events refer to intrinsic or qualitative properties are, in essence, questions about whether the events are of a type that has the right to play these two roles; here, it will be assumed that they are.

We represent the latter of these two ideas as ``Bell locality''.  To do this, we will need to define a \textit{full specification} of a region $\cA$.  This is an event $F$ that, when given, fixes the truth of every other event in $\cA$.  More precisely, it is an event belonging to $\cA$ such that, for every other event $X \in \alg(\cA)$, $F \subseteq X$ or $F \subseteq \bar{X}$ \cite{Henson:2005wb}.  The idea is that a full specification of the past of a region $\cA$ includes all events that might possibly influence events in $\cA$, and that any correlations between $\cA$ and a spacelike region $\cB$ should be a result of influences from these past events.  Bell gave a such a principle, calling it ``local causality'' (\cite{Bell:2004}, 1975, p.~54).

\paragraph{Bell locality:} for any two events $A$ and $B$ associated to spacelike-related regions $\cA$ and $\cB$ respectively, the following condition holds:
\begin{equation}
\label{e:so}
\mu(A|\lambda)\mu(B|\lambda)= \mu(A \cap B|\lambda) \quad \forall \lambda \in \Phi \bigl( \cP(\cA,\cB) \bigr),
\end{equation} 
where $\Phi \bigl( \cP(\cA,\cB) \bigr)$ is the set of full specifications of the region $\cP(\cA,\cB)$, which is a suitable past region for the pair $\{\cA,\cB\}$.

\paragraph{} Bell locality represents a conjunction of a principle of common cause (PCC) with relativistic causal structure.  Correlations between events should only be due to direct causal connections between the events, or due to some common cause, in which case they should disappear when we condition probabilities on all events that might be relevant to such a cause.  By adding relativistic causal structure, we ban the possibility of direct causal connection between spacelike events and ``all events that might be relevant'' to the common cause become a full specification of the past region (see \cite{Uffink:1999,Henson:2005wb} for more justification of this).

In the above, the definition of the past region has been left open deliberately.  It could be the intersection of the past lightcones of $\cA$ and $\cB$, the ``mutual past'', or their union (with $\cA$ and $\cB$ removed so that they do not lie in the past region), the ``joint past,'' or the past of one of the regions but not the other\footnote{Conditions of this type, with these different past regions, are shown to be equivalent in a similar framework in \cite{Henson:2005wb} (for some debate on whether the proof is sound, see \cite{Redei:2012,Henson:2012}).  However, the proof employed separability, called rule (\textit{iv}) there.}.  Below, we will see that the choice does not matter much for the issues at hand.

One of the conclusions becomes clear at this point: to define Bell locality, we have not had to assume separability.  We used nothing beyond the assumption that events happen in regions.  Separability is not implicit here at all, and the existence of non-separable events in the past region makes this definition no less reasonable.  They would be part of possible common causes in the past, and would be conditioned on as well. After conditioning the probabilities, there is no remaining correlation between $A$ and $B$.  That is all Bell's condition says.

To expand on this important point, imagine a laboratory, and an event that is associated to a region that extends for a light year around the lab in every direction at some moment in time (in the lab frame).  Locality implies that such an event should not affect anything in the lab for at least a year.  The point here is simply that this is true whether the event is separable or not.  The formal condition of Bell locality naturally incorporates this.   Under these circumstances it is clearly mistaken to argue, as some have, that the very existence of such an event would threaten our individuation of the lab system itself and thus any possible notion of locality (see sections \ref{s:einstein} and \ref{s:spekkens}).

Another point in the above discussion needs to be emphasised as an assumption, although this does not have such a formal character.

\paragraph{Operational consistency of localisation:}  Operational events are associated with the regions of spacetime in which they can be detected.

\paragraph{} Again, this condition does not imply separability.  It does not even imply that the operational events must be separable. But it is required to avoid absurd conclusions, as discussed at more length in section \ref{s:weakening}.

Going forward to discuss the EPRB experiment, it is good to take stock of what we are assuming.  We need to assume that some set of operational events are associated to regions, and that it makes sense to condition on a full specification of the past of those regions; although this follows from the localised events condition, which seems natural to impose, nothing more from this condition is strictly necessary.  We will also need to assume Bell locality, operational consistency of localisation, and a ``freedom of settings'' assumption.  We will \textit{not} assume separability in any form.

\subsection{The EPRB experiment}
\label{s:eprb}

In this experiment, we have two spacelike regions $\cA$ and $\cB$, representing two ``wings'' of our experimental apparatus.  In these regions some operational events are defined.  The events $A_s$ and $B_s$ correspond to whether one particular setting is chosen on some instrument in regions $\cA$ and $\cB$ respectively (by tradition, some orientation of a Stern-Gerlach magnet rather than another).  The events $A_o$ and $B_o$ refer to some outcomes (whether the spin is measured ``up'' according to the given orientation).  We will also refer to the variable $a_o$ that can take the value $A_o$ or $\bar{A_o}$, and so on\footnote{The are a number of choices in how to present the experiment.  Here I select the ``big history space approach'' \cite{Butterfield:2005} that includes $A_s$ and $B_s$ as events with probabilities of their own, rather than having to deal with the complication of treating them as parameters of a family of probability distributions which are nevertheless localised.  These doubtful probabilities could be removed without harming the argument in its essentials, as usual.  The experimental probabilities $\mu(a_o \cap b_o | a_s \cap b_s)$ do not depend on the absolute probabilities of the settings.}.

Applying Bell locality directly to events in $\cA$ and $\cB$, we have

\begin{equation}
\label{e:soeprb}
\mu(a_o \cap a_s \cap b_o \cap b_s | \lambda ) = \mu( a_o \cap a_s | \lambda ) \mu( b_o \cap b_s | \lambda ) ,
\end{equation} 
for all values of the variables and all full specifications of the past $\lambda$.  Here we introduce the freedom of settings assumption, essentially that the choice of settings is independent of past events:
\begin{equation}
\label{e:free}
\mu( a_s \cap b_s | \lambda ) = \mu( a_s \cap b_s ).
\end{equation} 
In this form, the condition makes the settings completely spontaneous in the sense that they depend on nothing at all in the past\footnote{This may not be so realistic; in a more general treatment, we might allow settings to correlate to some set of past events in the model which nonetheless were independent of every other event of relevance, or even allow limited correlations using some ``careful epsilonics'' as Bell puts it  (\cite{Bell:2004} 1977 p.102).  Another way to represent the same situation would be to replace the requirement that $\lambda$ be a ``full specification'' with the requirement that it has ``sufficient completeness for a certain accuracy'', as Bell explains  (\cite{Bell:2004} 1977 p.104).  See \cite{Norsen:2007, Seevinck:2011} for more on this. In any case, the main point of this article does not turn on the issues.}.

By summing over some variables, it follows from (\ref{e:soeprb}) and (\ref{e:free}) respectively that
\begin{equation}
\begin{split}
\label{e:blah}
\mu(a_s \cap b_s | \lambda ) &= \mu(a_s | \lambda) \mu(b_s | \lambda ) \\
&= \mu(a_s) \mu(b_s).
\end{split}
\end{equation} 
From this, we can see that
\begin{equation}
\mu(a_o \cap a_s \cap b_o \cap b_s | \lambda ) = \mu(a_o \cap b_o | a_s \cap b_s \cap \lambda ) \mu(a_s) \mu(b_s),
\end{equation}
and similar results are easily derived for each wing individually:
\begin{align}
\mu(a_o \cap a_s | \lambda ) &= \mu(a_o | a_s \lambda ) \mu(a_s), \\
\mu(b_o \cap b_s | \lambda ) &= \mu(b_o | b_s \lambda ) \mu(b_s).
\end{align}
Substituting these three results into (\ref{e:soeprb}) we have
\begin{equation}
\label{e:factorisability}
\mu(a_o \cap b_o | a_s \cap b_s \cap \lambda ) = \mu(a_o | a_s \cap \lambda ) \mu(b_o | b_s \cap \lambda ).
\end{equation} 
This last condition is commonly referred to as ``factorisability''.  The CHSH inequalities can be derived from it in a straightforward manner \cite{Clauser:1969a}.  This remaining part of the proof of Bell's theorem is essentially a series of elementary mathematical manipulations, and, at least in the published literature, it is uncontroversial that these steps make no use of separability.

What have we learned from this brief rehearsal of the well-known result?  Previously we saw that it is not necessary to assume separability in order to make sense of Bell locality, and that none of our assumptions imply separability.  Now we have seen that Bell inequalities can be derived in the normal manner without assuming separability.  That is, separability does not enter into the argument at all \footnote{\label{fn:healey}Healey once convincingly argued that a non-separable event in a region $\cA \cup \cB$ could be influenced by events to its past without violating ``relativistic locality'' \cite{Healey:1994}.  However, the conclusion that ``there is no sound reason'' to say that his interactive interpretation \cite{Healey:1989} of the EPRB experiment ``violates any defensible relativistic locality condition'' (\cite{Healey:1994}, p.~369) is unwarranted.  The argument above shows that it must violate one: the very condition that causes all the fuss about quantum non-locality.  It should be noted that Healey has since put forward a different interpretation, however \cite{Healey:2012}.} \footnote{Note also that the exact shape of the past region seemed not to enter, except insomuch as the region arguably has to be a certain shape to sufficiently specify the past (see appendix \ref{a:nouvelle}).  As long as freedom of settings (\ref{e:free}) refers to the same $\lambda$ as Bell locality (\ref{e:soeprb}) the proof will go through.}.

A reader versed in the lore of Bell's theorem might, at this point, wonder what has become of another condition sometimes labelled ``separability,'' Howard's ``separability of states''   \cite{Howard:1985} (\cite{Cushing:1989}, Howard pp.~224-253).  How do such conditions relate to the definition given here, and if they differ, is \textit{this} a hidden assumption of Bell's theorem?  These issues are discussed in appendix \ref{a:states} and section \ref{s:howard} respectively.

\subsection{Two criteria on which to judge locality arguments from considerations of signalling}
\label{s:signalling}

Before going on, it is important for the following discussion to note that Bell locality and freedom of settings together imply a prohibition on superluminal signalling, which will be referred to as ``no-signalling'' for short:

\begin{align}
\label{e:nosignalling1}
\mu(a_o | a_s \cap b_s) &= \mu(a_o | a_s), \\
\label{e:nosignalling2}
\mu(b_o | a_s \cap b_s) &= \mu(a_o | b_s).
\end{align}

It is well-known that, as predicted by QM, this condition holds in the EPRB experiment and in general, unlike the set of assumptions used to derive Bell's theorem.  It is relevant for later discussion to note a few basic consequences of this.

To start with, if one claims that a prohibition on superluminal influence is equivalent to a prohibition on superluminal signalling, there is no need to analyse Bell's result in order to save locality.  It is clear that, with this claim granted, locality is safe, and that is the end of the story.  However, Bell and many other commentators disagree with this claim, hence all the interest and literature surrounding Bell's theorem.  This argument will not be repeated in detail here, but is made by Bell (see \textit{e.g.} \cite{Bell:2004}, 1990, pp.~232-248) and has been defended by \textit{e.g.} Maudlin \cite{Maudlin:2002} and Norsen \cite{Norsen:2009}.  This means that, if we are worried about Bell's theorem at all, then we are worrying about a definition of superluminal influence that goes beyond superluminal signalling.  It would be one thing to take on the arguments of Bell directly, and to discuss in detail the difference between the two conditions and the motivation for Bell's locality.  However, it is another to simply re-assert the claim that no-signalling suffices.  Attempts of this type have already been answered in the earlier treatments.

Also, no-signalling can be thought of as a consistency check for the idea that Bell locality is a good representative of locality.  Bell locality implies no-signalling, given the freedom of settings assumption.  That assumption serves here to define a necessary feature of a genuine signal (that is, that the cause of the signal, the setting, is not correlated with the outcome only because of the influence of some past event).  This is necessary, because it is a foundational assumption that operational signalling is a form of influence, which we will not question here.

Both of these points will be relied upon in what follows to deal with some arguments that the assumptions of Bell's theorem could be weakened while preserving locality.  If the argument is just a round-about way of reducing locality to no-signalling, it will be dismissed (or more properly referred back to other writings on the subject including Bell's own).  If it leaves us with a candidate for locality that does not imply no-signalling by itself, it will be rejected as too weak.

\subsection{Weakening the assumptions?}
\label{s:weakening}

Separability has been shown to play no part in the derivation of Bell's inequalities.  However, considerations involving non-separability might give grounds to weaken the conditions set out above.  The aim would be to perform such a weakening while preserving a realisation of the locality principle.  Expanding on Teller's work, Morganti has made arguments broadly of this form \cite{Morganti:2009}.  We will examine similar ideas in this section.

Howard writes that ``[i]f two systems are not separable, then there can be no interaction between them, because they are not really \textit{two} systems at all'' (\cite{Howard:1985} p.~173, Howard's italics).  Similarly, Redhead writes that the particle in one wing of the EPRB experiment ``does not possess independent properties of its own'' (\cite{Redhead:1987}, p.107).  These comments give some inspiration for our task.   But they also beg the question of what in fact we are measuring and where it is, if statements of this sort are supposed to ease our minds on the problem of Bell's theorem.

To make this rather vague idea explicit, we need to translate it into a criticism of one of the assumptions made above.  Perhaps it is wrong to assume that an event $A$ associated to $\cA$ can only be influenced by events belonging to regions that lie \textit{entirely} in the causal past of $\cA$.  Perhaps it is only necessary, for non-separable events, that \textit{some} points of the region lie to the past of $\cA$\footnote{Morganti agrees with this proposal:
\begin{quote}
At most, the evidence requires one to put into doubt what, following
Jones, one may call ‘causal separability’, that is, the
requirement that an event $A$ can be the cause of another event $B$ only if
$A$ has a part entirely in the past light cone of $B$ that entirely causes $B$.
Indeed, if the emergent property of the whole system is such that it is
affected in its entirety by a measurement localized where one particle is
and, as a consequence of this, determines a new categorical property of
another particle at a different location, it is clear that causal separability
fails. But this, as explained, is not in itself a violation of locality.
\end{quote}
(\cite{Morganti:2009}, p.~1033).The reference is to \cite{Jones:1991}.  Morganti clearly states his belief that ``locality,'' which can be taken to share the meaning it is given here, can be preserved while making this move.}.  This motivates altering Bell locality to condition not only on a full specification of the past, but also on any non-separable events whose intrinsic regions overlap the past region.

However, if we take the past region $\cP$ for Bell locality to be the union of the pasts of $\cA$ and $\cB$, it is indeed possible that a full specification of the region could include non-separable events not contained in the past of $\cA$ alone.  But the proof of Bell's theorem still stands, as we have seen.  Perhaps then, not only can events be influenced by non-separable events associated to regions only partially in their past, but also, events can influence non-separable events associated to regions only partially in their future?  This does not work either: with this modification Bell locality no longer bans superluminal signals.   Consider two spacelike regions $\cA$ and $\cB$ as above, and a region $\cC$, some points of which lie to the future of $\cA$, and some points of which lie to the past of $\cB$.   If we accept the proposed modification of Bell locality, any events in $\cA$, even settings, can be correlated to events in $\cC$ and thus to $\cB$, without any restrictions.  So this proposed weakening clearly amounts to denial of the locality principle.

There is another way to challenge to the above framework, which might also be seen as a version of an argument that is present in the literature\footnote{ The following quote from Morganti illustrates this:
\begin{quote}
In other words, a measurement on an entangled system, commonly understood as an event $E_1$ localized where one of the particles is and determining another event $E_2$ localized where the other particle is in fact an event $E_1$ located everywhere the total system is (in particular, at the locations of both component particles) that determines events $E_2$ and $E_3$ localized at different places and yet in physical and spatiotemporal continuity with their cause, $E_1$. This means that one has a process that is entirely local at each stage.
\end{quote}
(\cite{Morganti:2009}, pp.~1031-1032).  This seems consistent with the version made explicit here, and the quotes from Howard and Redhead given earlier in this section also tend in this direction.}.  The idea is to weaken operational consistency of localisation.  For the EPRB setup, perhaps the outcome $A_o$ in the $\cA$ wing could be said to belong to $\cA \cup \cB$, and not to $\cA$ as before.  This seems to be what is suggested by saying that such events have no separate existence in the separate wings.  Note that, for this to undermine the proof of Bell's theorem, no operational event correlated to this outcome that happens subsequently could be assigned to a region using operational consistency of localisation, either.  For instance, if someone in the lab might remark ``look, the spin is up'' as a result of that outcome, and we assign \textit{this} event to $\cA$ (or any region spacelike to $\cB$), we can still derive a Bell inequality using \textit{this} event instead of $A_o$ -- even though the quantum entanglement has been eliminated by the time the person speaks.

Consider a situation in which an experimenter in the $\cA$ wing sees a light flash.  The correlation between this and the flashing of another distant light cannot be explained by some common cause in the past, according to our assumptions.  The suggestion above is that one could say to the worried experimenter ``don't worry, when you saw the flashing light, that actually corresponded to an event in the whole experimental region, not an event in your lab.  So you see, it's all local... Actually, your reaction to the flashing light didn't happen in your head either, but in the larger region.''  If we open the door to such arguments, a more elegant way to ensure locality would be to associate every event to one point, despite the convictions of the metaphysically na\"ive on the matter.  More specifically, it is in no way more unreasonable to apply this kind of reasoning to a hypothetical case of superluminal signalling than it is to apply it to outcomes in the EPRB experiment.  This, as has been repeatedly emphasised, clearly entails a violation of locality.

In both of these scenarios, it might seem that we can escape by falling back on an assumption of prohibition on superluminal signalling.  Is not \textit{that} what causes the unpalatable conclusions?  But if we rely on this, we may has well have avoided analysis of Bell's theorem by rejecting all locality assumptions except no-signalling in the first place.  Our policy, as set out in the previous section, is to dismiss such arguments in the context of analysing Bell's theorem.

Thus, we see that the various pieces of wordage on separability to be found in the literature, attractive though they might seem, start to look considerably less attractive in the light of the framework used above.  Once they are cast into an explicit form by pointing to a specific place at which the assumptions of Bell's theorem should be weakened, they can be overturned with elementary arguments.  There may be more weakenings of the assumptions of Bell's theorem that can be motivated by some words about non-separability.  To my knowledge, no other such option has been put forward with sufficient clarity to justify adding it to the list.   It is a strongly motivated conjecture that any other attempt of this sort could be just as easily ruled out.

\section{Previous writings on separability}
\label{s:people}

In the previous section, we saw that denying separability does not save us from the derivation of Bell inequalities.  The discussion there was, hopefully, fairly straightforward.  The greater difficulty comes in comparing this conclusion to previous writings no the subject, in order to see how misconceptions have arisen.

Since a number of the arguments surrounding non-separability and holism are motivated in large part by Jarrett's analysis of the EPRB experiment \cite{Jarrett:1984} (\cite{Cushing:1989}, Jarrett pp.~60-79), and use terminology related to it, it is worth taking a slight detour to examine these ideas, ahead of criticisms of Teller and Howard.

\subsection{Jarrett}
\label{s:jarrett}

The relevant part of Jarrett's analysis starts with the application of Bell causality to the EPRB set-up, (\ref{e:soeprb}) above.  Assuming freedom of settings, he resolved this condition into two which together imply (\ref{e:soeprb}).  The first he calls ``completeness'', which Shimony (\cite{Cushing:1989}, Shimony~pp. 25-37) renamed ``outcome independence'':
\begin{align}
\label{e:outcomeindep}
\mu(a_o | a_s \cap b_s \cap b_o \cap \lambda ) &= \mu(a_o | a_s \cap b_s \cap \lambda ), \\
\mu(b_o | a_s \cap b_s \cap a_o \cap \lambda ) &= \mu(b_o | a_s \cap b_s \cap \lambda ),
\end{align} 
while the second he called ``locality'' (``parameter independence'' for Shimony) which will be referred to here as Jarrett locality:
\begin{align}
\label{e:jarretloc1}
\mu(a_o | a_s \cap b_s \cap \lambda ) &= \mu(a_o | a_s \cap \lambda ), \\
\label{e:jarretloc2}
\mu(b_o | a_s \cap b_s \cap \lambda ) &= \mu(a_o | b_s \cap \lambda ).
\end{align}
Note that this relies on treating outcomes and settings differently, unlike Bell's local causality in which they are all ``beables''.  The only difference between them in the previous discussion was, naturally, in the freedom of settings condition, (\ref{e:free}).

This mathematical derivation is used as the pushing-off point for a certain view of Bell's theorem, given in a generic form here.  Jarrett emphasises the claim that Jarrett locality is equivalent to no-signalling, stating  ``LOCALITY $\Leftrightarrow$ NO SUPERLUMINAL SIGNALS'' in a box in one paper (\cite{Cushing:1989}, Jarrett pp.~69).  He holds that Bell locality is saying something stronger than this, to do with principles of common cause.  But he also refuses to view a failure of Bell locality as a failure of locality, relying on his own definition instead (in his argument he is chiefly concerned with potential conflict between QM and special relativity).  Howard, as we will see, goes a little further in his wording, identifying Jarrett locality with a satisfactory realisation of the locality principle.

Jarrett's analysis has been criticised on many grounds \cite{Maudlin:2002, Norsen:2009}.  Maudlin points out that a prohibition on signalling does not imply Jarrett locality, and so Jarrett locality should not be confounded with no-signalling as in eqn.~(\ref{e:nosignalling1}).  It is not hard to see that Jarrett locality, along with freedom of settings as in (\ref{e:free}), implies no-signalling.  But the converse is not true in general.  A number of authors have noted that in de Broglie-Bohm pilot wave mechanics, for instance, there is no-signalling in EPRB experiments, even though Jarrett locality is clearly violated \cite{Butterfield:1992a,Maudlin:2002, Jones:xxxx, Berkovitz:1998, Norsen:2009}.  The confusion comes from shifting the definition of $\lambda$.  When talking about no-signalling, Jarrett slips from treating past events $\lambda$ as some hypothetical variables describing the past which may not be operationally accessible (as in Bell's analysis), to treating them as a known initial state $s$ of an operational character, dropping consideration of any ``hidden'' past events.  The value of such an $s$ could be known to the agents, so that they can condition their probabilities on it, and then it is clear from (\ref{e:jarretloc1}) and  (\ref{e:jarretloc2}) that changing the value of a setting could change the measured probabilities for a spacelike outcome if Jarrett locality was abandoned.  But that is not necessarily the case when we take Bell's meaning for $\lambda$, as the pilot wave example makes clear.

We will add here that, were it not for the error in identifying Jarrett locality with no-signalling, Jarrett's argument could be taken as a review of Bell's theorem and no-signalling in the EPRB experiment.  He merely notes that no-signalling relieves us of the most obvious potential conflict with relativity.  So, were it not for this error, there would be nothing new in claiming that Jarrett locality is sufficient to ensure the locality principle (as Howard does).  This claim would merely be the assertion that locality is the same thing as no-signalling, against the warnings of Bell, as already considered in section \ref{s:signalling}.

\subsection{Teller}
\label{s:teller}

In his writings on Bell's theorem \cite{Teller:1986} (\cite{Cushing:1989}, Teller pp.~208-223), Teller is concerned with reconciling relativity with quantum mechanics.  He says he will ``call any theory which embodies these local actions and no superluminal propagation requirements Relativistic Causal theories'', that is, Relativistic Causality is for him (at least at this point in the paper) a combination of what are here called local action and locality.  He then claims that ``our unhesitating acceptance of relativistic causal theories... involves an assumption so basic to the thinking of most of us that we are not even aware that we are making it''.  This is the assumption of separability, which he calls an application of ``particularism'': ``[i]n application to relativistic theories, particularism takes the form of supposing the theory to apply exclusively to spacetime points and their non-relational properties'' (p.213).  This definition agrees with the separability principle above, and is adequately represented by the separability condition given in the framework of section \ref{s:framework}.  He goes on to add ``we should question the unspoken assumption, particularism, \textit{which sets the precondition for getting any relativistic causal theory off the ground}'' (my italics).  The following argument is offered later in the article as a justification of this claim:
\begin{quote}
But when (or insofar as) particularism is denied, the idea of causal locality has no application.  Causal locality concerns the lawlike connection between nonrelational properties applying (in this discussion) to space-time points.  To say that causal locality has been violated most plausibly should be taken to mean that there are nonrelational properties of space-time points which are related in some other way (lawlike dependencies) at a distance or through superluminal causal chains.  On the other hand, when we are concerned with nonsupervening relations, this circle of ideas has no grip.
\end{quote}
 (\cite{Cushing:1989}, Teller p.~215).

This is little more than a reiteration of the claim.  It is supposedly implausible to attribute a violation of local causality to properties that do not apply at spacetime points, but no explanation is given for this.  Instead, Teller again begs the question by assuming that ``causal locality'' is only plausible as a principle when applied to a separable theory.  We have seen that this is not the case in the previous section.

Later in the same article Teller makes the argument more specific for the probabilistic version of Bell's theorem reviewed above.  He gives a version of Jarrett's conditions and argues, with Jarrett, that outcome dependence is the part of Bell locality that we must reject.  The problem still remains, however, of explaining outcome dependence as an instance of failure of particularism rather than anything suggestive of a conflict with relativity.  To this end, Teller states a version of the principle of common cause due to Lincoln Moses, claims that ``Moses was unwittingly presupposing particularism,'' and adds the supposed assumption explicitly to the PCC.  Adding relativistic causal structure, Teller gives his weakened version of what is here called locality: ``when particularism holds, relativistic causal theories exclude superluminal causal propagation.''  He then argues that ``if one grants the assumptions about a particularist world, failure of outcome independence must involve some kind of failure of particularism''.

Let us set aside the fact that, in the same article, relativistic causality was previously \textit{defined} to mean local action and no superluminal propagation, but the argument is now that relativistic causality only implies these things under the condition of particularism, giving the impression that this attractive-sounding principle could be saved if one denies particularism.  There are larger problems.  Teller's argument depends on the assumption that locality, and indeed the PCC itself, only make sense under the assumption of separability (``particularism''), and otherwise do not.  This point is returned to often in the article, and is depended on at the conclusion.  But there is no substantial argument to justify this assumption anywhere the article.  The passages quoted above are the most explicit to be found.   To make matters worse this claim is false, as has been demonstrated in section \ref{s:framework}.  Without such a justification, what remains is the form of argument referred to in the introduction: adding an unnecessary assumption to Bell's theorem and then claiming to have saved locality by removing it.   One can only suppose that the incorrect and unsubstantiated assertion is implicitly taken to have been established in the literature already, perhaps from an interpretation of Howard's work.


\subsection{Howard and Howard's Einstein}
\label{s:howard}

Howard also wrote on separability, using a certain reading of Einstein as inspiration for a treatment of Bell's theorem  \cite{Howard:1985} (\cite{Cushing:1989}, Howard pp.~224-253).  His position is distinct from Teller's in several significant ways.  In the earlier of the two articles, he writes that ``separability says that spatially separated systems possess separate real states''.  In his later, more careful definition, Howard's separability means that disjoint regions are such that ``(1) each possesses its own, distinct physical state, and (2) the joint state of the two systems is wholly determined by these separate states.''  While (at least in the light of the discussion above) the first part might suggest something like the localised events principle, the second is the part that has led to definitions of separability such as that used above.  In the same place, Howard distinguishes two kinds of separability:

\begin{quote}
The more modest concerns the individuation of states; it is the claim that spatio-temporally separated \textit{systems} do not always possess separable \textit{states}, that under certain circumstances either there are no separate states or the joint state is not completely determined by the separate states.  I call this way of denying the separability principle the \textit{non-separability of states}.  The more radical denial may be called the \textit{non-separability of systems}; it is the claim that spatio-temporal separation is not a sufficient condition for individuating systems themselves, that under certain circumstances the contents of two spatio-temporally separated regions constitute a single system.
\end{quote}
(\cite{Cushing:1989}, Howard p.~226, Howard's italics).

Howard's inspiration comes from a number of quotes from Einstein, primarily the following, which precedes an argument for the incompleteness of orthodox quantum mechanics:

\begin{quote}
If one asks what is characteristic of the realm of physical ideas independently of the quantum-theory,
then above all the following attracts our attention: the concepts of physics refer to a real external
world, i.e., ideas are posited of things that claim a “real existence” independent of the perceiving
subject (bodies, fields, etc.), and these ideas are, on the one hand, brought into as secure a relationship
as possible with sense impressions. Moreover, it is characteristic of these physical things that they
are conceived of as being arranged in a space-time continuum. Further, it appears to be essential for
this arrangement of the things introduced in physics that, at a specific time, these things claim an
existence independent of one another, insofar as these things ``lie in different parts of space.'' Without
such an assumption of the mutually independent existence (the ``being-thus'') of spatially distant
things, an assumption which originates in everyday thought, physical thought in the sense familiar
to us would not be possible. Nor does one see how physical laws could be formulated and tested
without such a clean separation. Field theory has carried this principle to the extreme, in that it localizes within infinitely small (four-dimensional) space-elements the elementary things existing independently of one another that it takes as basic, as well as the elementary laws it postulates for them.

For the relative independence of spatially distant things (A and B), this idea is characteristic: an external influence on A has no \textit{immediate} effect on B; this is known as the ``principle of local action'', which is applied consistently only in field theory.  The complete suspension of this basic principle would make impossible the idea of the existence of (quasi-) closed systems and, thereby, the establishment of empiricaly testable laws in the sense familiar to us.
\end{quote}
(\cite{Howard:1985} p.~187-188, translated from \cite{Einstein:1948} pp.~321-322.)

In the following section this quote will be discussed further, but for now the focus will be on Howard's reading of it.  From this quote, Howard claims that ``Einstein's `principle of local action' and his `assumption of the mutually independent existence of spatially distant things' correspond, respectively, to the locality and separability principles.'' (\cite{Cushing:1989}, Howard p.234).  In other words, Howard's claim is that Einstein here is describing two different principles, separability in the first paragraph and locality in the second.  He goes on to claim that separability and locality are independent assumptions of Bell's theorem.

Whatever Howard's ``non-separability of systems'' is, ``non-separability of states'' is more clearly modeled on the ``non-separability'' apparent for entangled quantum states.  Howard explains that ``at a minimum, the idea is that no information is contained in the joint state that is not already contained in the separate states, or, alternatively, that no measurement result could be predicted on the basis of the joint state that could not already be predicted on the basis of the separate states'' (\cite{Cushing:1989}, Howard p.~226).  This, along with the above definition, will be called the ``separability of states principle'' in the following.   In the same place, Howard defines a ``locality principle'', saying that ``the relativistic version of the principle asserts that a system's state is unaffected by events in regions of spacetime separated from it by a spacelike interval'' (\cite{Cushing:1989}, Howard p.~227).  This is sufficiently similar to the locality principle stated above as to be identified with it for the purposes of this argument.

For the EPRB experiment, Howard goes on, ``I define a state as a conditional probability measure assigning probabilities to outcomes conditional upon the presence of global measurement contexts''.  He then formalises separability of states as a specific condition in the case of the EPRB set-up by making the following identifications:
\begin{align}
\label{e:howard1}
\mu(a_o | a_s \cap b_s \cap \lambda) &= \mu_\alpha(a_o | a_s \cap b_s), \\
\label{e:howard2}
\mu(b_o | a_s \cap b_s \cap \lambda) &= \mu_\beta(b_o | a_s \cap b_s),
\end{align}
``where $\alpha$ and $\beta$ represent the separate states of the systems in the $\cA$ and $\cB$ wings, and $\lambda$ represents the joint state''\footnote{\label{fn:laudisa}Laudisa also criticises Howard's arguments on separability, and his main criticism turns on this point \cite{Laudisa:1995}.  He argues that, if $\alpha$ and $\beta$ are defined as representing these separate states, then the identifications (\ref{e:howard1}) and (\ref{e:howard2}) are not justified.  For instance, in Bohm's theory we might be tempted on this basis to make $\alpha$ just the position of the particle headed to the $\cA$ wing and similarly for $\beta$.  But $\lambda$ should also contain the quantum state in that case, as that is part of the ontology as well, and so (\ref{e:howard1}) and (\ref{e:howard2}) would not necessarily hold in that case.  However, the association of the wave-function to a region has been left ambiguous here: there is nothing inconsistent with Bohm's theory in including a copy of the quantum state at, say, every point in space (an observation made by Lucien Hardy, private communication), and so it could be in both $\alpha$ and $\beta$.  The arguments against Howard's views given here are different from Laudisa's: for instance, Laudisa does not challenge Jarrett's views on Bell's theorem or Howard's interpretation of Einstein.  Related, though distinct, arguments are made in appendix \ref{a:states}.  For now, we treat the equations simply as identifications that define the meaning of $\alpha$ and $\beta$, accepting Howard's definitions, and assess the consequences of this.}.  For Howard, $\lambda$ is taken to be the joint state of the two wings before the other relevant events  (in Howard's notation $\lambda$ is also a label on probability distributions).  The separability of states condition is then
\begin{equation}
\label{e:howardsep}
\mu(a_o \cap b_o | a_s \cap b_s \cap \lambda ) = \mu_\alpha(a_o | a_s \cap b_s) \mu_\beta(b_o | a_s \cap b_s).
\end{equation} 
With the above definitions, this ``separability of states'' turns out to be equivalent to Jarrett's ``completeness'' as in eqn.~(\ref{e:outcomeindep}).  Thus it is a consequence of the assumptions of Bell's theorem.  Now, by this point, the definition of separability of states has diverged from the definition of separability given in section \ref{s:locevents}, and arguably Howard's own original definition, but discussion of this point will be relegated to appendix \ref{a:states} in order not to distract from a more important problem with Howard's argument.

This is the basis on which Howard argues that separability and locality are independent assumptions of Bell's theorem.  After repeating Jarrett's erroneous claim that Jarrett locality is equivalent to no-signalling, he explicitly takes Jarrett locality as in eqn.~(\ref{e:jarretloc1}) to be an adequate representative of the (relativistic) locality principle set out above: ``[o]n Jarrett's analysis, a violation of the Bell inequality need \textit{not} entail relativistic nonlocality, because it may result \textit{either} from a violation of the Jarrett locality condition, \textit{or} from a violation of his completeness condition'' (\cite{Cushing:1989}, Howard p.230, Howard's italics).

It has already been argued in section \ref{s:jarrett} that removing the reliance on Jarrett would clarify such arguments.  What remains after this is done is nothing but a claim that no-signalling can be identified with locality, and stronger conditions of locality (such as Bell's) are therefore unwarranted.  This entails a disagreement with Bell's original arguments for preferring Bell locality, as noted in section \ref{s:signalling}.  Howard's argument must boil down to some new reason to reject Bell's arguments for local causality, if it is to have any content.

Simply rejecting the formal condition of ``separability of states'' as in eqn.~(\ref{e:howardsep}) (which is implied by Bell locality) is not therefore the crux of the argument.  The crux is, rather, the claim that ``separability of states'' is not an implication of the locality principle, properly understood.  Why not?  Howard simply asserts that Jarrett locality has more right to the name of locality, as it is equivalent to no-signalling.  He does not take on Bell's justifications for Bell locality directly.  Thus there is no new argument against Bell's claim that no-signalling is not enough. Howard's argument has no new content once Jarrett's erroneous claim is removed.

It remains to consider what Einstein and Bell wrote about separability and its relation to their arguments about locality. 

\subsection{Einstein}
\label{s:einstein}

In this paper a number of different notions of ``independence of events associated to disjoint/spacelike regions'' have been distinguished and some of their consequences have been discussed.  A number of quite different concepts have been revealed, which are, however, often referred to with the same words.  This raises questions over Howard's interpretation of Einstein.  In contrast to Howard's reading of the quoted passages, it is consistent to interpret Einstein's claim that ``things claim an existence independent of one another'' with something along the lines of the localised events condition as given in section \ref{s:framework}, or at least, something weaker than the separability principle.  That is, to each region of spacetime is attributed certain events in an unambiguous manner.  Dropping this more basic assumption, rather than separability, gives much more reason to state that ``without such an assumption... physical thought in the sense familiar to us would not be possible.''  

When Einstein says that ``[f]ield theory has carried this principle to the extreme'' he gives a definition more in line with the separability principle used here: ``it localizes within infinitely small (four-dimensional) space-elements the elementary things existing independently of one another that it takes as basic.'' If this is the ``extreme'' case and not the general case that Einstein explains earlier in the paragraph, this surely implies that what he said earlier in the paragraph is \textit{not} equivalent to separability but something weaker.

If we accept Howard's argument, then we must attribute to Einstein an inability to imagine physical thought without separability.  Howard provides a possible reason for this.  For Howard, Einstein is arguing that ``spatio-temporal separation is the only conceivable objective criterion of individuation'' of systems, which may be necessary for any conceivable physics  (\cite{Cushing:1989}, Howard p.241).  Whether this is correct or not, we can now ask: is not the localised events assumption a good enough basis on which to individuate systems?  This principle supplies a definition of what is associated to a region $\cX$ and what is not (any non-separable events associated to regions overlapping but not contained in $\cX$ are not), which should be sufficient\footnote{Healey recognises this when he writes, referring to the same quote from Einstein, that ``[e]ach of $A$ and $B$ may be spatially localised and have its own state, even if the state of the nonlocalised $AB$ does not supervene on the states of $A$ and $B$''(\cite{Healey:1989}, p.352).}.  Thus, most of Howard's speculations about Einstein's motivations for assuming separability, arguably, actually apply to the assumption of localised events only (Einstein's enthusiasm for field theories is also mentioned by Howard as a motivation, and might be taken to apply only to separability, but this carries little weight when divorced from the larger argument).  If what Einstein is calling for is simply the existence of localised events, the problem of having to attribute arbitrary prejudices to Einstein is considerably ameliorated. 

This ambiguity shows through in other places, such as the following:
\begin{quote}
...we might therefore all along have been testing not simply local hidden variable theories, but separable, local hidden variable theories.

I suspect that most of our trouble in understanding so-called `quantum nonlocality' is a result of this more basic confusion.  We focus our attention on the apparent demonstration of non-local effects mysteriously communicated between two systems separated by a spacelike interval, without pausing to ask the deeper question of whether there are really \textit{two} systems, or just \textit{one}.  We can and should clear up this confusion.
\end{quote}
(\cite{Howard:1985} p.~195-196, Howard's italics).

Here Howard identifies the proposition that the two wings of the EPRB experiment can safely be treated as two systems, in some unspecified sense, with the separability principle.  It is only through a series of unjustified identifications of this sort that we arrive, from Einstein's quote, to Howard's separability of states.  Without these problematic identifications it is impossible to get from one to the other.

There is a strong case, therefore, that Einstein was \textit{not} grounding his discussion of the incompleteness of QM with the separability principle\footnote{He does say that \textit{local action} is ``applied consistently only in field theory,'' which might seem to undermine the argument (originally from Healy \cite{Healey:2007}) that local action and separability are not linked, but a more plausible reading is that field theory was the only type of theory available at the time of writing in which local action is unambiguously enforced.}.

If we suitably define ``separability'' then of course it may be identified with what Einstein is calling for.  In that case identifying the content of Einstein's statements with ``separability of states'' as in eqn.~(\ref{e:howardsep}) is the step of Howard's argument that requires justification.  But if it is consistent to read Einstein as calling only for localised events, then any such justification will fail: the existence of localised events clearly is not equivalent to Howard's separability of states.

The main claim in this article has been that separability is not a prerequisite for defining locality, and neither is it an assumption of Bell's theorem.  Here, the point is that Einstein did not write that the separability principle was a prerequisite for his argument that quantum mechanics is incomplete, which inspired Bell.  He only claims that real things may be ``brought into as secure a relationship as possible with sense impressions'' and that ``these things claim an existence independent of one another, insofar as these things `lie in different parts of space.' ''  Though not as formal as the conditions set out above, these statements could be taken to rule out such moves as rejecting operational consistency of localisation, or the principle of localised events.  The intention could be that any such move would lead to absurdities of the type discussed in section \ref{s:weakening}, but Einstein is not specific about that here.

In conclusion, there is a strong argument that the ``localised events'' interpretation of Einstein's statement should be preferred to Howard's ``separability'' interpretation. At the very least, when we keep in mind this possible alternative interpretation, there is nothing in any quotes Howard uses that unambiguously goes further than this on separability and its relation to the completeness of quantum mechanics. In this light it seems like an unwarranted over-interpretation to say that Einstein was arguing for separability in any strong sense, such as the separability principle used here.


\subsection{Bell, Spekkens and Harrigan}
\label{s:spekkens}

A remaining question is whether Bell intended separability to be thought of as a grounding assumption of his locality condition and of his theorem, as is claimed in the article by Harrigan and Spekkens \cite{Harrigan:2010}.  Their definition of separability is as follows.
\begin{quote}
Suppose a region $R$ can be divided into local regions $R_1,R_2,...,R_n$.  An ontological model is said to be \textit{separable} only if the ontic state space $\Lambda_R$ of region $R$ is the Cartesian product of the ontic state spaces $\Lambda_{R_i}$.''
\end{quote}
(\cite{Harrigan:2010} p.9).  The ontic state space here plays a similar role to $\Omega$ above, with the difference that this ``state space'' refers to states at one time rather than histories.  Apart from this, the definition is essentially the same as that given in section \ref{s:locevents}\footnote{This definition is at the level of the state/history space whereas, in section \ref{s:locevents}, the slightly higher level of the algebra of events was used.  In this article, the set of all events $\Sigma(\cM)$ can be \textit{any} $\sigma$-algebra on the history space, and so a condition like Harrigan and Spekkens' would not be appropriate.  But if $\Omega$ is countable and $\Sigma(\cM)$ is fixed to be the Boolean algebra of subsets of $\Omega$, the definitions coincide.}.  They add the following as justification of their claim that separability is ``a necessary component of any sensible notion of locality'':
\begin{quote}
The assumption of separability is made, for instance, by Bell when he restricts his attention to theories of \textit{local} beables.  These are variables parameterising the ontic state space `which (unlike for the total energy) can be assigned to some bounded space-time region' ''.
\end{quote}
The quote within the quote is from Bell (\cite{Bell:2004}, 1976, p.53).  This claim has little effect on their illuminating discussion of Einstein's arguments for the incompleteness of quantum mechanics and its relation to the epistemic view of quantum states.  It does however have some relevance to future directions.  They comment at the end of their paper that ``ontological models that are fundamentally relational might also fail to be captured by the framework described here.''  This seems to leave the door open to the possibility that non-separable or ``holistic'' models could escape from the conceptual problems of Bell's theorem by coming up with a ``sensible notion of locality'' that does not rely on separability.  It should be noted, however, that Spekkens has recently expressed a different view, arguing that Bell did not in fact need to assume separability for his theorem, whilst still maintaining that Bell's locality condition, as he stated it, presupposes separability \cite{Spekkens:2012}.

Let us analyse the claim in \cite{Harrigan:2010} in the light of the definitions given in section \ref{s:framework}.  The quote from Bell only talks of being able to assign spacetime regions to beables, \textit{i.e.}~it only justifies the assumption of the existence of localised events.  It says nothing about whether events belonging to regions can be resolved into combinations of events in smaller regions.  Neither is there anything else to suggest that he is making such an assumption in the article referred to.  It is more likely that this idea stems from Howard's interpretation of Einstein and Bell's theorem, which is referenced as well.

All this does not absolutely rule out the possibility that Bell intended separability to be thought of as an assumption of his theorem.  However, as Spekkens notes in his later treatment \cite{Spekkens:2012}, the following quote suggests that he did not.

\begin{quote}
It is notable that in this argument nothing is said of the locality, or even localizability, of the variable $\lambda$ [which has the same meaning as above].  These variables could well include, for example, quantum mechanical state vectors, which have no particular localization in ordinary space-time.  It is assumed only that the outputs $A$ and $B$ [$a_o$ and $b_o$ here], and the particular inputs $a$ and $b$  [$a_s$ and $b_s$ here], are well localized. 
\end{quote}
(Bell 1981, p.~153)\footnote{Thanks to Michel Buck for pointing out this quote.}.  In the first sentence quoted here, Bell seems to be deliberately repudiating the idea that separability is an assumption of his theorem.  How else to read ``localizability'' when explicitly opposed to locality, especially when illustrated by the example of including the quantum state as a beable?  Furthermore, the assumptions referred to certainly do not amount to any claim about separability: they are exactly calling for the inputs and outputs to be localised events\footnote{Note that Bell does need his past variables to be associated to \textit{some} past region, even if an infinite one.  Therefore he is not arguing against the existence of localised events in the weakest sense here, although the \textit{spatial} localisation of the wave-function might most naturally be an infinite region, which seems almost equivalent to saying it is not localised anywhere.}.

There seems to be nothing anywhere else in Bell's published writings on the subject to suggest that he believed separability to be important for his theorem.  In some of the earliest writings, Bell uses the word ``separability'' to refer to quantum states (\cite{Bell:2004}, 1966, p.9), and he states that ``It is the requirement of locality... that creates the essential difficulty.  There have been attempts to show that even without such a separability or locality requirement no `hidden variable' interpretation of quantum mechanics is possible''(\cite{Bell:2004}, 1964, p.14).  Here he seems to be using the word interchangeably with locality, referring to only \textit{one} previously mentioned principle.  After this he seems to drop the use of the term altogether.  At several other points he unambiguously speaks of the existence of localised ``beables'', as in the quote used by Spekkens and Harrigan, but not about the need to assume separability, which is distinct as has already been argued.

This illustrates Bell's genius for focusing on essentials: if he had included separability as a condition, this would have been unnecessary, as we have seen.

\section{Conclusion: the challenge of reconciling locality and the quantum}

Hopefully the arguments in this article have bought some clarity to claims about non-separability in the context of Bell's theorem.  The main points that have been argued for are summarised here.  There is a weaker principle, localised events, which merely states that all physical events can be associated to spacetime regions in a consistent manner.  This principle does not imply separability.  The definition of Bell's locality condition does not rely on separability in any way.  The proof of Bell's theorem does not use separability as an assumption.  If,  inspired by considerations of non-separability, the assumptions of Bell's theorem are weakened, what remains no longer embodies the locality principle.  Teller's argument for relational holism as a solution to the problem of Bell's theorem relies on the unjustified assumption that separability underpins Bell locality.  Howard's argument that separability of states should be dropped is, once reliance on Jarrett is identified and excised, only a call to give up Bell's notion of locality in favour of a ban on superluminal signalling, something that has already been criticised in the literature.  His claim that Einstein grounded his arguments on non-locality and the incompleteness of QM with the assumption of separability is highly questionable; the quotes from Einstein are better interpreted as referring to the assumption of localised events, or something similar.  Finally, in the only places that he could be interpreted as treating this issue, Bell only endorses localised events, and at one point seems to explicitly reject the idea that separability is an assumption of his theorem.

The moral of this story goes beyond the case of non-separability.  Serious claims that locality (when that word takes the meaning it does here) can be reconciled with QM are well worth careful investigation, as argued in the introduction.  If correct, this would be an important achievement and could lead to great advancement in our understanding of QM, possibly with momentous consequences.  However, any such claim must be able to survive rigorous interrogation.  As Bell says before defining local causality,
\begin{quote}
Now it is precisely in cleaning up intuitive ideas that one is likely to throw the baby out with the bathwater.  So the next step should be viewed with the utmost suspicion...
\end{quote}
(\cite{Bell:2004}, 1990, p.239).

In the same spirit, this is a good attitude to take towards any attempt to replace the definition he then gave.  In particular, the following questions must be answered before any such approach can be considered as even plausible.  Firstly, what definition of the locality principle is being mooted, and is it stated as unambiguously as Bell's?  Secondly, is the principle strong enough to ban superluminal signalling, as Bell's does when combined with freedom of settings?  Is it also \textit{more} than merely a ban on superluminal signalling, plausibly satisfying Bell's motivations in defining local causality?  Can it give the right answers for other uncontroversial examples of locality and non-locality (\textit{e.g.}~an agent or robot rolling some dice and sending the results to many distant locations, or a comet travelling at twice the speed of light relative to the Earth)?  Also (something that was not relevant above but nevertheless may scupper other approaches), does it continue to hold good after past events are conditioned on, avoiding ``Simpson's paradox'' \cite{Uffink:1999,Henson:2005wb}?

There are many other further directions for research suggested by the arguments given in this paper.  Now that relevant terms have been clarified and some results derived, advocates the idea of non-separability could move the debate forward by engaging with this position in its specifics.  On another point, Howard's interpretation of Einstein has been highly influential, but it has been challenged here.  This brief challenge could be expanded into a more comprehensive reinterpretation of Einstein's views on the subject.  For instance, what was Einstein's attitude towards the principle of local action, when it is distinguished from locality and separability as above?  Another fairly obvious extension would be to apply the same style of discussion to other approaches to reconciling locality and QM.  For example, there are a number of approaches that, broadly speaking, rely on denying reality to the measurement outcome in one wing of the EPRB experiment \cite{Brown:2002a, Smerlak:2006gi, Fuchs:2010, Healey:2012}\footnote{In \cite{Brown:2002a} Brown and Timpson make a careful distinction between causal principles based on the PCC and what they define as locality, and so their approach is anything but na\"ive to the sort of questions being raised here.  Such reasoning could be applied to the other approaches listed here as well.  Nevertheless they all explicitly attempt to reconcile locality with quantum mechanics, using definitions of locality similar to that employed here.} while others argue that the common cause principle could be implemented in a weaker way \cite{Hofer:1999, Hofer:2002} (see \cite{Uffink:1999,Henson:2005wb,Butterfield:2007a,Henson:2012} for some arguments already in the literature that tend to undermine the latter position).  It is not clear from current expositions whether these attempts to save locality can survive a comparison to the minimal requirements listed above.

A clear view of the difficulties met with when applying these requirements to non-separability, and other approaches, helps to motivate the consideration of novel approaches that can tackle these concerns directly.  For instance, new possibilities arise when we introduce a new modality dealing with \textit{indefiniteness}, allowing us to preface standard propositions with ``it is definite that...'' and  ``it is indefinite that...'' (this is similar to Reichenbach's idea of introducing a third truth value \cite{Reichenbach:1944a}, although with different rules and motivations).  This introduces a number of new choices on how to define causal influences, and thus locality.  In forthcoming work, it will be claimed that at least one of these choices allows quantum correlations while satisfying the minimal requirements listed above, offering an interesting way forward that arguably satisfies some of both Einstein's and Bohr's intuitions \cite{Henson:2013a}.

If this point is conceded, the outstanding question will be whether this argument really provides a satisfying alternative to Bell's conditions that can do all (or any) of the work we expect a causal principle to do.  Constructing simple models that employ this idea, and reproduce interesting features of quantum mechanics, might be a reasonable next step here.  All this would be part of the ongoing task of restoring a meaningful, and useful, picture of the microworld to modern physics.

\section*{Acknowledgments}

The author would like to thank Michel Buck for pointing out a useful quote from Bell, Fay Dowker and Rafael Sorkin for many conversations about causality and relativity, and especially to Harvey Brown for pointing to previous work on separability such as that of Healey.  Thanks are also due to Travis Norsen and Rob Spekkens for edifying discussions of an earlier version on the manuscript.  This work was made possible through the support of a grant from the John Templeton Foundation.

\begin{appendix}

\section{A complication: local action and \textit{nouvelle} locality}
\label{a:nouvelle}

In section \ref{s:framework} it was argued that Bell locality can be formulated without first assuming separability.  However, there is another definition of Bell's ``local causality'' given in a later article (\cite{Bell:2004}, 1990, pp.~232-248).  It is reasonable to ask: it possible that, in order to use \textit{this} definition of locality, we must assume separability?  And would that be problematic for the main theses of this paper?

In this case, the past region is not taken to be whole causal past of any of the relevant regions.  Instead we are asked to consider a ``slice'' of spacetime (a region between two spacelike hypersurfaces).  Now, equation (\ref{e:so}) must hold for the past region consisting of ``at least those parts of [the slice] blocking the two backward lightcones,'' that is, the intersection of the slice with the union of the causal pasts of $\cA$ and $\cB$.  We will call this region $\cS$ and this version of the principle ``\textit{nouvelle} locality''.  Bell writes that ``what happens in the backwards lightcone of [the regions in question]'' should be ``sufficiently specified, \textit{for example} by a full specification of local beables in [a slice of the past light cone]' '(\cite{Bell:2004}, 1990, p.240, my italics).  This suggests that Bell intended the condition to hold for \textit{any} slice as long as we have reason to think that it ``sufficiently specifies'' the past.  But what does, and how do we know it does?

Although \textit{nouvelle} locality might look like a stronger condition than Bell locality, it is possible for correlations between events in  $\cA$ and $\cB$ to be \textit{introduced} rather than removed by conditioning on more past events, leading to the so-called ``Simpson's paradox'' \cite{Uffink:1999,Henson:2005wb}; this is essentially the reason why we must ``sufficiently specify'' the past.  What if such a ``Simpson's paradox'' arose due to past events outside of $\cS$?  These include events associated to regions to the past of $\cS$, as well as non-separable events lying partially to the past and partially to the future of it. It is reasonable to demand that our causal principle rules this out for the same reason as it should rule out correlations after conditioning on events in $\cS$.

From this is seems that conditioning on Bell's ``example,'' the region $\cS$, only sufficiently specifies the past under certain conditions.

\paragraph{Strong relativistic local action (SRLA):} for any two events $X$ and $Y$ belonging to regions $\cX$ and $\cY$ such that $\cX$ lies entirely to the past of $\cY$, the following condition holds:
\begin{equation}
\label{e:srla}
\mu(X|\lambda)\mu(Y|\lambda)= \mu(X \cap Y|\lambda) \quad \forall \lambda \in \Phi \bigl( \cD(\cA,\cB) \bigr),
\end{equation} 
where $\Phi \bigl( \cD(\cX,\cY) \bigr)$ is the set of full specifications of the region $\cD(\cX,\cY)$ which is the intersection of some slice with $J^+(\cX) \cap  J^-(\cY) \backslash (\cX \cup \cY)$\footnote{This condition is called ``strong'' as it is possible to imagine weaker version of the condition that set $\cD$ to be the whole of $J^+(\cX) \cap  J^-(\cX)$ or a slightly smaller region.  However such versions would not save \textit{nouvelle} locality, as they would still allow causal propagation via non-separable events which might span $\cS$.}.

\paragraph{} This rule seems strong enough to prevent influences from propagating from some past event to $\cA$ or $\cB$ without also showing up in $\cS$.  I conjecture that, assuming SRLA for both cases, Bell locality and \textit{nouvelle} locality will turn out to be equivalent\footnote{Something similar is probably true of other examples in which the whole past is not conditioned on, like Percival and Penrose's causal condition in which the past region is a ``wedge'' such that the remainder of $J^+(\cA) \cup  J^-(\cB)$ when the wedge is removed is a disjoint union of two regions, one containing $\cA$ and the other $\cB$  \cite{Penrose:1962}.}.   A proof could conceivably run along roughly the same lines as those given in \cite{Henson:2005wb} for other variations on the shape of the past region.  Without the SRLA assumption, conditioning only on events in a slice is problematic: they may not sufficiently specify the past, as Bell required.

It seems unlikely that Bell changed the shape of the past region with the intention of introducing an implicit local action assumption.  Instead, the real intention of the change may have been to relieve worries expressed elsewhere.  Bell refers to ``the necessity of paying attention in such a study to the creation of the world'' in a different context (\cite{Bell:2004}, 1977, p.~102) and then, in a footnote, adds that ``The invocation in [(\cite{Bell:2004}, 1975, pp.~52-62)] of a \textit{complete} account of the overlap of the backward light cones is embarrassing in a related way, whether going back indefinitely or to a finite creation time... In more careful discussion the notion of completeness [full specification] should perhaps be replaced by that of sufficient completeness for a certain accuracy, with suitable epsilonics'' (\cite{Bell:2004}, 1977, p.104).  With \textit{nouvelle} locality nothing arbitrarily far back into the past need be conditioned on, resolving this possible worry in a cleaner way.  But there seems nothing wrong with Bell's original pragmatic suggestion if we do not wish to make this move.

There is relevance to the discussion of separability.   The condition rules out correlations due to a common cause amongst non-separable events that span $\cS$, or a correlation mediated to $\cA$ and $\cB$ by non-separable events that span $\cS$.  This does not imply separability, but it does mean that it is not necessary to specify these non-separable events in order to screen off spacelike events.  This threatens to trivialise their dynamics.  But note that events associated to some subset of a spacelike hypersurface in $\cS$ do not face this problem because they cannot span $\cS$.  So non-separable events of this kind  cause no more problems when conditioning only on a slice.  In one way the opposite is true: Bell does not rule out the extension of the past region to the whole of a slice in the above definition, in which case the specification could indeed include things like the quantum state which have no obvious ``locality, or even localizability'' in space at all\footnote{This is also of relevance to Healey's thesis that locality and local action can both be preserved in gauge field theory only if separability is sacrificed \cite{Healey:2007}.  The events there are defined on closed spacelike curves.  It would be interesting to formulate local action and locality as specifically as is done here, and to explicitly show that they are not violated for certain gauge theories in Healey's picture of them.}.

In any case, the version of Bell locality used in section \ref{s:belllocality} arguably is sensible despite Bell's ``embarrassment''.  It is not obviously a problem to condition on events arbitrarily far into the past, because events do not have to be empirically accessible to be conditioned on in a model, as argued in \cite{Henson:2012}.  The point of Bell's theorem is that we cannot find a Bell local model that gives the same correlations as quantum mechanics \textit{no matter what} we hypothesise about past events in our model.  Also, locality and local action are different principles, and any condition that enforces locality need not also enforce local action.

In conclusion, whichever version of Bell's condition we prefer, separability as defined in section \ref{s:locevents} need not be assumed in order to make the condition sensible, and so the main arguments in the paper are not threatened when this alternative formulation of Bell locality is considered.

\section{Separability vs. Howard's separability of states}
\label{a:states}

In this appendix, we will discuss Howard's formalisation of separability of states, given in section \ref{s:howard}, and find the connection to the definition of separability in section \ref{s:locevents}.  The relevant definitions are (\ref{e:howard1}) and (\ref{e:howard2}) and the condition itself, eqn.~(\ref{e:howardsep}).

It is useful to note at the outset that the definitions of $\mu_\alpha(a_o | a_s \cap b_s)$ and $\mu_\beta(b_o | a_s \cap b_s)$ do not imply much about the meaning of the ``states'' $\alpha$ and $\beta$ \cite{Laudisa:1995}.  For example, consider a model of the EPRB experiment in which all the past events are associated to one point.  In this case $\alpha$ and $\beta$ clearly do not label initial states of two spatially disjoint systems corresponding to the two wings.  However, it is possible to add an extra assumption that the relevant past region $\cP$ is the disjoint union of two regions, $\cP_\cA$ and $\cP_\cB$, which lie to the past of the $\cA$ and $\cB$ wings respectively (this would accord with local action and \textit{nouvelle} locality as in the preceeding appendix).

Even making this assumption, it is easy to see that separability does not imply eqn.~(\ref{e:howardsep}).  Take for example a theory with correlations between the outcomes, but a past with no non-trivial events, separable or otherwise (or any model that gives these probabilities after conditioning on some separable past events).  Maudlin gives a similar argument, adding to the picture some tachyons to make it more intuitive and ``mechanical'' (maintaining local action for instance) (\cite{Maudlin:2002}, pp.~97-98)\footnote{Other models can be imagined: for example, a spherical wave propagating out from one outcome that affects the other, or a string between a pair of entangled particles along which local actions could pass.   If the question is what chance there is of formulating a natural and empirically adequate theory that manifests outcome dependence and separability, thoughts of this kind might be of interest (see \cite{Berkovitz:1998} for a discussion); the question at hand, however, is whether separability implies separability of states over all theories, and so these considerations are not relevant.  In this sense, Maudlin's argument against Howard is not reliant on any ill-defined ``intuitions about tachyons'' \cite{Berkovitz:1998}.}.

To put a finer point on this, consider for example a ``back-yard'' EPRB-like experiment, which can be modelled as above, but in which the wings are \textit{not} required to be spacelike.  Let us imagine that one outcome $A_o$ (random, according to the theory we apply) triggers a ping-pong ball to be fired into the $\cB$ wing.  By some simple mechanism, this affects the outcome there.  To complete the argument, note that Howard's original definitions of separability say nothing about whether the ``regions of spacetime'' concerned are spacelike or timelike to each other.  Now, this back-yard experiment violates Howard's formal condition (\ref{e:howardsep}) as applied to its ``wings,'' but obviously this simple scenario does not rely on violating Howard's ``fundamental ontological principle,'' separability.   It follows that the formal separability of states condition does not faithfully express a ban on non-separable events overlapping the two wings, the idea that the two wings are in fact two separate systems, the claim that basic properties are all associated to points, lack of mysterious``passion-at-a-distance'' or anything of the sort.  There is nothing motivating the condition -- unless that is, we put the wings spacelike to each other, and invoke \textit{locality} in order to prevent such scenarios.  We already know that Howard's separability of states condition follows from Bell's locality condition, and it is best interpreted simply as a consequence of that well-founded condition.

Conversely eqn.~(\ref{e:howardsep}) does not imply separability, or even the relevant consequence of separability: that the are no events associated to $\cP_\cA \cup \cP_\cB$ that are not logical combinations of events associated to $\cP_\cA$ and $\cP_\cB$.  The equation means nothing more or less than it explicitly says: that $\mu(a_o \cap b_o | a_s \cap b_s \cap \lambda )$, where $\lambda$ is the state of  $\cP$, \textit{which may well have associated to it non-separable events}, equals $\mu_\alpha(a_o | a_s \cap b_s) \mu_\beta(b_o | a_s \cap b_s)$.

So much for the comparison itself. As in the main text, the more difficult question to answer is why the conclusions reached in this framework differ from those already in the literature.  There is an implicit assumption in Howard's treatment that resolves this problem.  Howard and those who have developed his argument \cite{Winsberg:2003,Fogel:2007} must keep the meaning of $\lambda$ that Bell gave, in order to keep their arguments about separability of states in contact with the derivation of the Bell inequalities.  Now, $\lambda$, for Bell, ranges over all \textit{full specifications} of $\cP$ (see section \ref{s:belllocality}).  However, Howard formally defines his ``states'' $\lambda$ as $p_\lambda(x|m)$, which translates to $\mu(a_o \cap b_o | a_s \cap b_s \cap \lambda )$ here  (\cite{Cushing:1989}, Howard p.~226).  The two definitions are only consistent if the following assumption holds: $\mu(a_o \cap b_o | a_s \cap b_s \cap \lambda )$ completely determines the full specification $\lambda$, and similarly for $\alpha$ and $\beta$.  This assumption is not part of the framework used in this paper, which is one reason for the discrepancy in conclusions.  More importantly, the assumption is problematic.  For example, if we add a third setting value, requiring the new condition for $\mu_\alpha(a_o | a_s \cap b_s)$ does not imply the same condition for its restriction back to two settings.  In other words, the argument goes through when our theory only allows our apparatus to have two marks on its dial, but not if we imagine a third unused mark\footnote{This argument is closely related to Norsen's criticism of Jarrett's interpretation of Bell's theorem, which also depends on the definition of $\lambda$, although Norsen's account goes into more depth.  It can also be usefully compared to Laudisa's criticism of Howard \cite{Laudisa:1995}.}!

Even given this assumption, there is another problem.  Winsberg and Fine also argue that separability does not imply (\ref{e:howardsep}), but on the grounds that, to satisfy Howard's principle, the joint ``state'' could determine the ``states'' of the wings \cite{Winsberg:2003} in \textit{any} way, not necessarily through a product (see also Fogel's detailed account \cite{Fogel:2007}).  With Howard's implicit assumption, Winsberg and Fine's weaker condition has some justification: if, as separability implies, the full specifications $\alpha$ and $\beta$ determine $\lambda$, then the assumption implies that $\mu_\alpha(a_o | a_s \cap b_s)$ and $\mu_\beta(b_o | a_s \cap b_s)$ will determine $\mu(a_o \cap b_o | a_s \cap b_s \cap \lambda )$ (Fogel's ``functionwise'' composition).  However, when Winsberg and Fine conclude that both locality and separability can be preserved in the wake of Bell's theorem, they are relying on the rest of Howard's (and Jarrett's) arguments, including this implicit, and flawed, assumption.

In conclusion, we saw in sections \ref{s:howard} and \ref{s:einstein} that (\textit{a}) even if we allow the definition of separability of states to go unquestioned, Howard's argument that locality can be saved by dropping this condition is flawed, and (\textit{b}) the attribution of the separability assumption to Einstein is questionable.  We can now add that (\textit{c}) the definition in eqn.~(\ref{e:howardsep}) is not a good formalisation of the separability principle as Howard defines and discusses it.

This analysis is relevant to the so-called PBR theorem, which rules out the epistemic interpretation of quantum states under a number of conditions, including a so-called ``preparation independence'' condition \cite{Pusey:2012} that bears some similarity to a separability condition.  As the debate on this theorem and its conditions goes on, care must be taken to not to read too much into this condition.  In this case, however, the line between epistemological and ontological concepts has been made clear by the original authors, giving hope that the discussion will remain clear-cut on this issue.

\end{appendix}

\bibliographystyle{h-physrev3}
\bibliography{refs}

\begin{thebibliography}{10}

\bibitem{Bell:2004}
J.~Bell,
\newblock {\em Speakable and unspeakable in quantum mechanics}, second ed.
  (Cambridge Univeristy Press, Cambridge, 2004),
\newblock Subsequent references to Bell's papers will be made in-line with the
  year of publication and the page numbers from this book. The original
  publication details can be found therein.

\bibitem{Redhead:1987}
M.~Redhead,
\newblock {\em Incompleteness, nonlocality and realism} (Oxford University
  Press, Oxford, 1987).

\bibitem{Cushing:1989}
J.~T. Cushing and E.~McMullin, editors,
\newblock {\em Philosophical Consequences of Quantum Theory} (Notre Dame Press,
  Notre Dame, Indiana, 1989),
\newblock Subsequent references to papers in this collection will be made
  in-line with author's name and the page numbers from this book.

\bibitem{Maudlin:2002}
T.~Maudlin,
\newblock {\em Quantum Non-Locality and Relativity}, second ed. (Malden,
  Massachusetts, 2002).

\bibitem{Norsen:2006a}
T.~{Norsen},
\newblock Foundations of Physics Letters {\bf 19}, 633 (2006),
  arXiv:quant-ph/0601205.

\bibitem{Norsen:2007}
T.~Norsen,
\newblock Arxiv preprint , 19 (2007).

\bibitem{Norsen:2007a}
T.~{Norsen},
\newblock Foundations of Physics {\bf 37}, 311 (2007), arXiv:quant-ph/0607057.

\bibitem{Norsen:2009}
T.~Norsen,
\newblock Foundations of Physics {\bf 39}, 273 (2009), arXiv:0808.2178.

\bibitem{Butterfield:1992a}
J.~Butterfield,
\newblock British Journal for the Philosophy of Science {\bf 43}, 41 (1992),
  pss/687884.

\bibitem{Brown:2002a}
C.~G. {Timpson} and H.~R. {Brown},
\newblock Entanglement and relativity,
\newblock in {\em Understanding Physical Knowledge}, edited by R.~Lupacchini
  and V.~Fano, Department of Philosophy, University of Bologna, 2002,
  quant-ph/0212140.

\bibitem{Henson:2013a}
J.~Henson,
\newblock {Causality, Bell's theorem, and ontic definiteness},
\newblock In preparation.

\bibitem{Reichenbach:1956a}
H.~Reichenbach,
\newblock  (Berkeley: University of California Press, 1956),
\newblock reissued 1991.

\bibitem{Jarrett:1984}
J.~Jarrett,
\newblock {No{\^u}s} {\bf 18}, 569 (1984).

\bibitem{Wessels:1985}
L.~Wessels,
\newblock No{\^u}s {\bf 19}, 481 (1985).

\bibitem{maudlin:2010}
T.~Maudlin,
\newblock American Journal of Physics {\bf 78}, 121 (2010).

\bibitem{Howard:1985}
D.~Howard,
\newblock Stud. Hist. Phil. Sci {\bf 16}, 171 (1985).

\bibitem{Teller:1986}
P.~Teller,
\newblock British Journal for the Philosophy of Science {\bf 37}, 71 (1986).

\bibitem{Healey:1989}
R.~Healey,
\newblock {\em The Philosophy of Quantum Mechanics} (Cambridge Univeristy
  Press, Cambridge, 1989).

\bibitem{Healey:1991}
R.~A. Healey,
\newblock The Journal of Philosophy {\bf 88}, 393 (1991).

\bibitem{Healey:1992}
R.~Healey,
\newblock Philosophical Topics {\bf 20}, 181 (1992).

\bibitem{Healey:1994}
R.~Healey,
\newblock Stud. Hist. Philos. Sci. A {\bf 25}, 337 (1994).

\bibitem{Morganti:2009}
M.~Morganti,
\newblock Philosophy of Science {\bf 76}, 1027 (2009).

\bibitem{Laudisa:1995}
F.~Laudisa,
\newblock The British Journal for the Philosophy of Science {\bf 46}, 309
  (1995).

\bibitem{Berkovitz:1998}
J.~Berkovitz,
\newblock Stud. Hist. Philos. Mod. Phys. {\bf 29}, 183  (1998).

\bibitem{Winsberg:2003}
E.~Winsberg and A.~Fine,
\newblock Journal of Philosophy. {\bf C}, 80–97 (2003).

\bibitem{Fogel:2007}
B.~Fogel,
\newblock Stud. Hist. Philos. Mod. Phys. {\bf 38}, 920  (2007).

\bibitem{Healey:2007}
R.~Healey,
\newblock {\em Gauging what's real} (Oxford University Press, New York, USA,
  2007).

\bibitem{Butterfield:2005}
J.~Butterfield,
\newblock British Journal for the Philosophy of Science {\bf 57}, 709 (2005).

\bibitem{Spekkens:2004a}
R.~W. Spekkens,
\newblock Phys. Rev. {\bf A75}, 032110 (2007), arxiv:quant-ph/0401052.

\bibitem{Harrigan:2010}
N.~Harrigan and R.~W. Spekkens,
\newblock Foundations of Physics {\bf 40}, 125 (2010), arXiv:0706.2661.

\bibitem{Bartlett:2012}
S.~D. Bartlett, T.~Rudolph, and R.~W. Spekkens,
\newblock Phys. Rev. A {\bf 86}, 012103 (2012), arXiv:1111.5057.

\bibitem{Pusey:2012}
M.~F. Pusey, J.~Barrett, and T.~Rudolph,
\newblock Nature Phys. {\bf 8}, 476 (2012), arXiv:1111.3328.

\bibitem{Spekkens:talk2012}
R.~Spekkens,
\newblock Why {I} am not a psi-ontologist,
\newblock Seminar given at Perimeter Institute, Waterloo ON, Canada, 2012,
  http://pirsa.org/12050021/.

\bibitem{Spekkens:2012}
R.~Spekkens,
\newblock {FQXi} essay contest, spring  (2012), arXiv:1209.0023.

\bibitem{Ghirardi:1994}
G.~Ghirardi and R.~Grassi,
\newblock Studies in History and Philosophy of Science {\bf 25}, 397 (1994).

\bibitem{Henson:2005wb}
J.~Henson,
\newblock Stud. Hist. Philos. Mod. Phys. {\bf 36}, 519 (2005),
  quant-ph/0410051.

\bibitem{Uffink:1999}
J.~Uffink,
\newblock Philosophy of Science {\bf 66}, 525 (1999).

\bibitem{Redei:2012}
M.~{R{\'e}dei} and I.~{San Pedro},
\newblock Stud. Hist. Philos. Mod. Phys. {\bf 43}, 84  (2012), arxiv:1204.4288.

\bibitem{Henson:2012}
J.~Henson,
\newblock Stud. Hist. Philos. Mod. Phys. {\bf 44}, 17  (2013), arXiv:1210.1463.

\bibitem{Seevinck:2011}
M.~Seevinck and J.~Uffink,
\newblock Not throwing out the baby with the bathwater: Bell’s condition of
  local causality mathematically ‘sharp and clean’,
\newblock in {\em Explanation, Prediction, and Confirmation}, edited by
  D.~Dieks, W.~J. Gonzalez, S.~Hartmann, T.~Uebel, and M.~Weber Vol.~2, pp.
  425--450, Springer Netherlands, 2011, arXiv:1007.3724.

\bibitem{Clauser:1969a}
J.~Clauser, M.~Horne, A.~Shimony, and R.~Holt,
\newblock Phys. Rev. Lett. {\bf 26}, 880 (1969).

\bibitem{Healey:2012}
R.~Healey,
\newblock British Journal for the Philosophy of Science {\bf 63}, 729 (2012),
  arXiv:1008.3896.

\bibitem{Jones:1991}
M.~R. Jones,
\newblock {\em Locality and Holism: The Metaphysics of Quantum Theory},
\newblock PhD thesis, Stanford University, Stanford, CA, 1991.

\bibitem{Jones:xxxx}
M.~Jones and R.~K. Clifton,
\newblock Against experimental metaphysics,
\newblock in {\em Midwest Studies in Philosophy 18}, edited by P.~French,
  T.~Wehling, and H.~Wettstein, pp. 295--316, University of Notre Dame Press,
  Notre Dame.

\bibitem{Einstein:1948}
A.~Einstein,
\newblock Dialectica {\bf 2}, 320 (1948).

\bibitem{Reichenbach:1944a}
H.~Reichenbach,
\newblock  (University of California Press, 1944),
\newblock Reprinted by Dover 1998.

\bibitem{Smerlak:2006gi}
M.~Smerlak and C.~Rovelli,
\newblock Found.Phys. {\bf 37}, 427 (2007), quant-ph/0604064.

\bibitem{Fuchs:2010}
C.~A. {Fuchs},
\newblock (2010), arxiv:1003.5209.

\bibitem{Hofer:1999}
G.~Hofer-Szab{\'o}, M.~{R{\'e}dei}, and L.~Szab{\'o},
\newblock British Journal for the Philosophy of Science {\bf 50}, 377 (1999).

\bibitem{Hofer:2002}
G.~Hofer-Szab{\'o}, M.~{R{\'e}dei}, and L.~Szab{\'o},
\newblock Philosophy of Science {\bf 69}, 623 (2002).

\bibitem{Butterfield:2007a}
J.~Butterfield,
\newblock The British Journal for the Philosophy of Science {\bf 58}, 805
  (2007), arXiv:0708.2192.

\bibitem{Penrose:1962}
R.~Penrose and I.~Percival,
\newblock Proceedings of the Physical Society {\bf 79}, 605 (1962).

\end{thebibliography}

\end{document}